\documentstyle[twocolumn,seceq,epsfig]{jpsj}

\def\runtitle{Molecular dynamics simulation of aging in amorphous silica}
\def\runauthor{Holger {\sc Wahlen} and Heiko {\sc Rieger}}

\title
{
Molecular dynamics simulation of aging in amorphous silica
}

\author
{ 
Holger {\sc Wahlen}$^1$ and Heiko {\sc Rieger}$^2$
}

\inst
{
$^1$John von Neumann Institute for Computing,
Forschungszentrum J\"ulich, 52425 J\"ulich, Germany\\
$^2$ FB 10.1 Theoretische Physik, Universit\"at des Saarlandes,
66041 Saarbr\"ucken, Germany
}

\recdate
{
\today
}

\newcommand{\tw}{t_{\rm w}}
\newcommand{\tr}{t_{\rm r}}
\abst
{
  By means of molecular dynamics simulations we examine the aging
  process of a strong glass former, a silica melt modeled by the BKS
  potential. The system is quenched from a temperature above to one
  below the critical temperature, and the potential energy and the
  scattering function $C (\tw, t + \tw)$ for various waiting times
  $\tw$ after the quench are measured. We find that both qualitatively
  and quantitatively the results agree well with the ones found in
  similar simulations of a fragile glass former, a Lennard-Jones
  liquid.
}

\begin{document}
\sloppy
\maketitle

\section{Introduction}

In the recent years aging effects have become a very active field of
research in the study of non-equilibrium systems. Common features have
been found in the aging behavior of spin and structural glasses
in experiments and simulations, and similarities in the theories
worked out indicate the possibility of a connection between the slow
dynamics in both types of glasses \cite{review}.

Even though silica (SiO$_2$), a strong glass former, has been examined
frequently both in theories and experiments, comparatively few
simulations have been performed so far on the dynamics of systems like
this that form an open network structure. In this paper we present
several results of aging simulations for silica and compare them to
those obtained for a model for a fragile glass former, a binary
Lennard-Jones mixture.

\section{Simulation}

\subsection{BKS potential}
\newcommand{\rjk}{r_{jk}}

The model we used in our simulations is given by the potential proposed
by van Beest, Kramer and van Santen (BKS)\cite{prl64-1955}:
\[
U_{jk} (\rjk) =
\frac{\gamma q_j q_k}{\rjk}
+ A_{jk} \exp (- B_{jk} \rjk) - \frac{C_{jk}}{\rjk^6}
\]
with $\gamma = e^2 / (4 \pi \varepsilon_0)$; the values of the
parameters, taken from the original publication, are listed in tables
\ref{tab:constants} and \ref{tab:ions}. As has been shown in previous
calculations and simulations, this potential reproduces many of both
the structural\cite{prb46-5933} and dynamical properties of silica and
has frequently been used for simulations of the amorphous phase.
Moreover, it has the advantage of consisting of two-body interactions
only, so that it can be implemented in a simulation more efficiently
than some other potentials proposed for silica that also contain
three-body terms. Considering that silica forms an open network where
the silicon atoms are the centers of tetrahedra and the oxygen atoms
work as bridges between them, it is not obvious at all that a
potential without three-body interactions is suitable for a good
description of the properties; it has been found, however, that the
competing two-body forces indeed mimic the three-body ones, causing
the formation of a network.\cite{prb54-15808}

Previous simulations\cite{prb60-3169} have shown that the critical
temperature for this model is $T_{\rm c} = 3\,330$~K, which is a good
approximation for the value found in experiments with real
silica, 3\,221~K\cite{hess}.

\begin{table}[tbh]
  \caption{Interaction constants in the BKS potential.}
  \label{tab:constants}
\begin{tabular}{@{\hspace{\tabcolsep}\extracolsep{\fill}}cccc}
$j$--$k$ & $A_{jk}$ [eV] & $B_{jk}$ [\AA$^{-1}$] & $C_{jk}$ [eV \AA$^6$] \\
\hline
O--O   &  1388.7730 & 2.76000 & 175.0000 \\
O--Si  & 18003.7572 & 4.87318 & 133.5381 \\
Si--Si &          0 &       0 &        0
\end{tabular}
\end{table}
\vskip-1.5cm
\begin{table}
  \caption{Charges and masses of the ions.}
  \label{tab:ions}
    \begin{tabular}{@{\hspace{\tabcolsep}\extracolsep{\fill}}ccc}
    & $q_j$  & $m_j$ [u] \\
\hline
O   & $-1.2$ & 15.9940 \\
Si  & $+2.4$ & 28.0855
    \end{tabular}
\end{table}

For the O--O and O--Si interactions, the BKS potential has the
drawback that it shows an unphysical behavior for small distances by
diverging towards minus infinity.  To correct this, it was replaced by
a harmonic potential for distances below the location of the potential
maximum (1.4387~\AA\ for O--O, 1.1936~\AA\ for O--Si). We confirmed
the observations made in earlier simulations\cite{prb54-15808}
already: because of the potential barrier the particles have to
overcome in order to reach them, these small distances only occur in
very few particle pairs, so the substitution of the potential is not
likely to have a relevant effect on the quantitative results of the
simulation.

\subsection{Parameter choices}

We performed molecular dynamics simulations in a cubic system of
length $L = 25.11$~\AA\ containing 1089 (363+726) particles, so that
the density was fixed at 2.3~g/cm$^3$. To integrate the equations of
motion, the velocity form of the Verlet algorithm with time steps of
1.6~fs was used. The non-Coulombic part of the potential was truncated
and shifted at a distance of 5.5~\AA. The Coulomb term was calculated
by means of the Ewald summation\cite{rapaport}, splitting it into a
sum in real space,
\[
U' = \gamma \sum_{j < k}
\frac{q_j q_k {\rm erfc} (\alpha \rjk)}{\rjk}
\]
(where erfc denotes the complementary error function, ${\rm erfc} (x)
= 2 \pi^{-1/2} \int_x^\infty \exp (- t^2) \, {\rm d}t$), and one
in Fourier space,
\[
U'' =
\frac{\gamma}{2 \pi L} \sum_{{\mib n} \neq 0} \frac{1}{n^2}
\exp \left(
  - \frac{\pi^2 n^2}{\alpha^2 L^2}
\right) \left|
  \sum_j q_j \exp \left(
    \frac{2 \pi {\rm i}}{L} {\mib n} {\mib r}_j
  \right)
\right|^2
\]
($n \equiv |{\mib n}|$). We chose $\alpha = 6.5/L$; the real-space sum
was truncated and shifted at 9~\AA, the Fourier-space one at $|{\mib
  n}| = 5$. Further details about the simulation can be found
elsewhere\cite{hw}.

\subsection{Preparation and measurements}

The simulations were done in the same way as previous ones for a
Lennard-Jones potential\cite{prl78-4581,prl81-930}: a system is first
equilibrated at a temperature above the critical one, then at $t = 0$
suddenly ``quenched'' to one below; there it is allowed to relax for
various waiting times before measurements of the potential energy and
the generalized scattering function
\[
C (\tw, \tw + t) = \frac{1}{N} \sum_j \exp ( {\rm i} {\mib
  q} \cdot [ {\mib r}_j (t + \tw) - {\mib r}_j (\tw) ] ).
\]
(where $N$ is the number of particles and $\mib q$ is a wave vector)
are started, so that examples for both an equilibrium constant and a
two-time correlation function are examined.

To improve the statistics of the results, we averaged over three
independent initial configurations consisting of randomly distributed
particles. Each of those was used to produce equilibrium
configurations at 6\,000, 7\,000 and 8\,000~K by first rescaling the
particle velocities periodically according to the desired temperature
and then letting the system evolve freely for 55\,000 time steps.

Each of the resulting configurations was then used for measurement
runs at 2\,700, 3\,000 and 3\,200~K: at $t = 0$, the system was
brought to the final temperature by rescaling the velocities, and the
temperature was controlled throughout the run by repeating this every
50 time steps. After waiting 0, 10, 100, 500, 1\,000, 3\,000, 10\,000
and 30\,000 steps, respectively, measurements of the potential
energy and the generalized scattering function were started and
performed every ten steps; the scattering function was calculated by
averaging over fifty randomly chosen wave vectors of absolute value
1.7~\AA$^{-1}$, the first sharp maximum of the structure factor for
silica\cite{prb60-3169}. These runs were executed on eight processors
of a Cray T3E with a time limit of four hours per run, which allowed
between 60\,000 and 75\,000 time steps; with the chosen step size,
this corresponds to 96--120 ps ($10^{-12}$sec.) in reality.

\section{Results}

\subsection{Potential energy}

\begin{figure}[htb]
  \caption{Potential energy of the system (shifted by 13\,900~eV)
    as a function of the time after the quench from 7\,000 to
    3\,000~K, plotted (a) linearly and (b, dotted) logarithmically.
    The dashed line in (b) shows the approximation for long times using
    a power-law dependence with an exponent of --0.22.}
  \label{fig:e730}
  \epsfig{figure=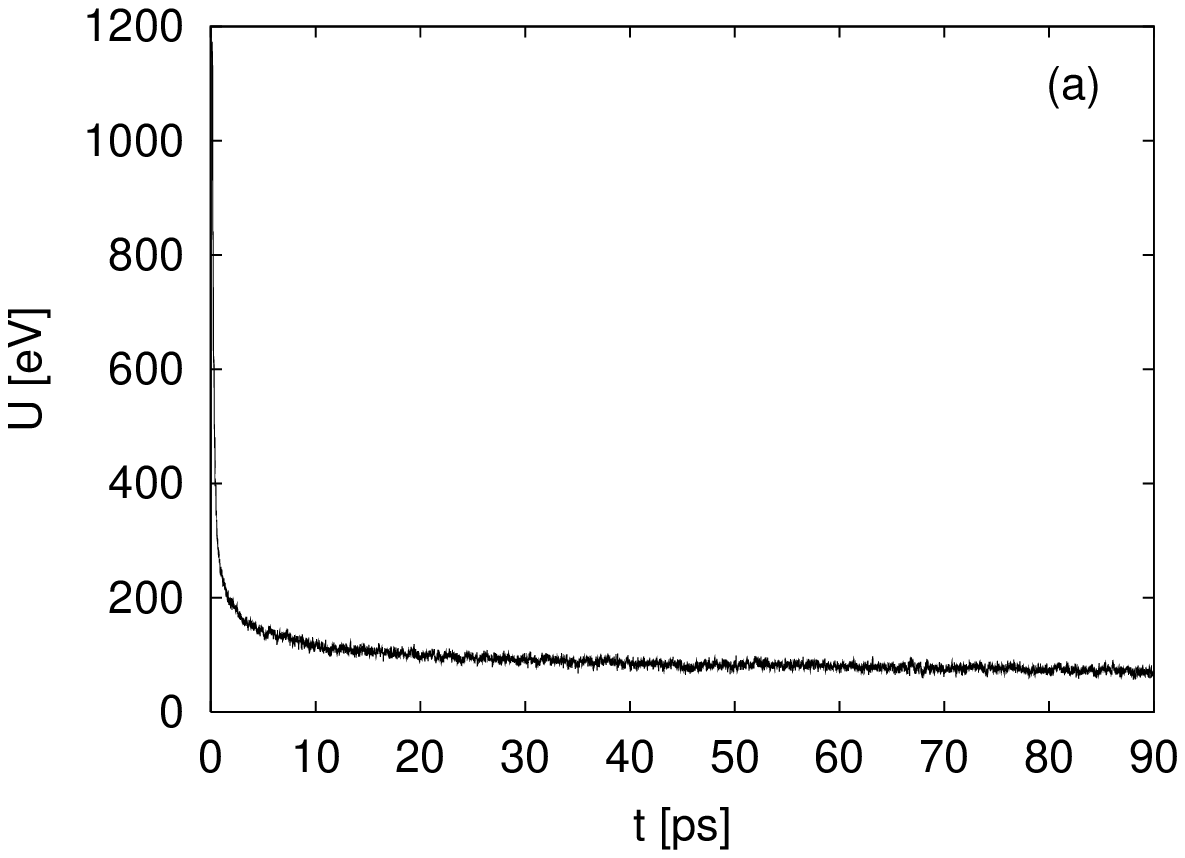,width=9cm}
  \epsfig{figure=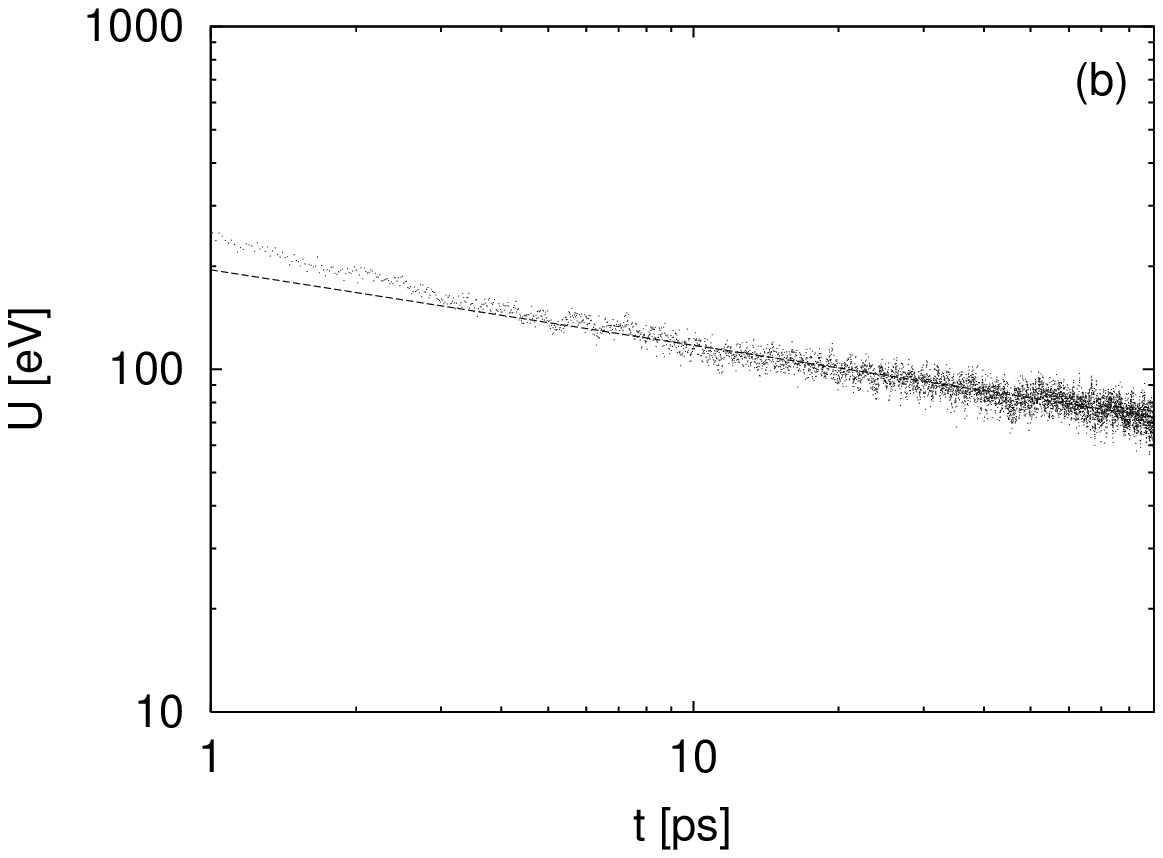,width=9cm}
\end{figure}

The potential energy shows a fast decay only for a short time after
the quench; after that it appears to remain constant at a first glance
in a linear plot (an example is shown in figure \ref{fig:e730}a), but
a closer examination by means of a logarithmic plot (figure
\ref{fig:e730}b) reveals that it still decays slowly.

It is possible to approximate the decay for long times by a power law.
The exponents for this approximation are listed in table
\ref{tab:pot-exponents}; the average value for all quenches is
--0.20.

\begin{table}[htb]
    \caption{Estimated exponents for the time dependence of the
      potential energy for the various combinations of initial (rows)
      and final (columns) temperatures.}
    \begin{tabular}{@{\hspace{\tabcolsep}\extracolsep{\fill}}c|ccc}
& 2\,700~K & 3\,000~K & 3\,200~K \\
\hline
6\,000~K & --0.22 & --0.18 & --0.15 \\
7\,000~K & --0.26 & --0.22 & --0.18 \\
8\,000~K & --0.28 & --0.18 & --0.16
    \end{tabular}
    \label{tab:pot-exponents}
\end{table}

Next we compare these results to those obtained in analogous
simulations for a binary Lennard-Jones (LJ) system. This model glass
consists of a binary mixture of spheres which interact via an LJ
potential with a strong $r^{12}$ repulsive and a weak $r^6$ attracting
part (van der Waals force); the spheres have two different radii to
avoid crystallization. The mixture is a so called fragile glass former
that has a glass transition at around $T=0.46$ in the appropriate
LJ-units.

\begin{figure}[htb]
  \caption{Generalized scattering function $C (\tw, \tw + t)$ as a
    function of the time for the quenches from 7\,000~K (top) and
    8\,000~K (bottom) to 3\,000~K and various selected waiting times
    $t_{\rm w}$.}  \label{fig:gsf}
    \epsfig{figure=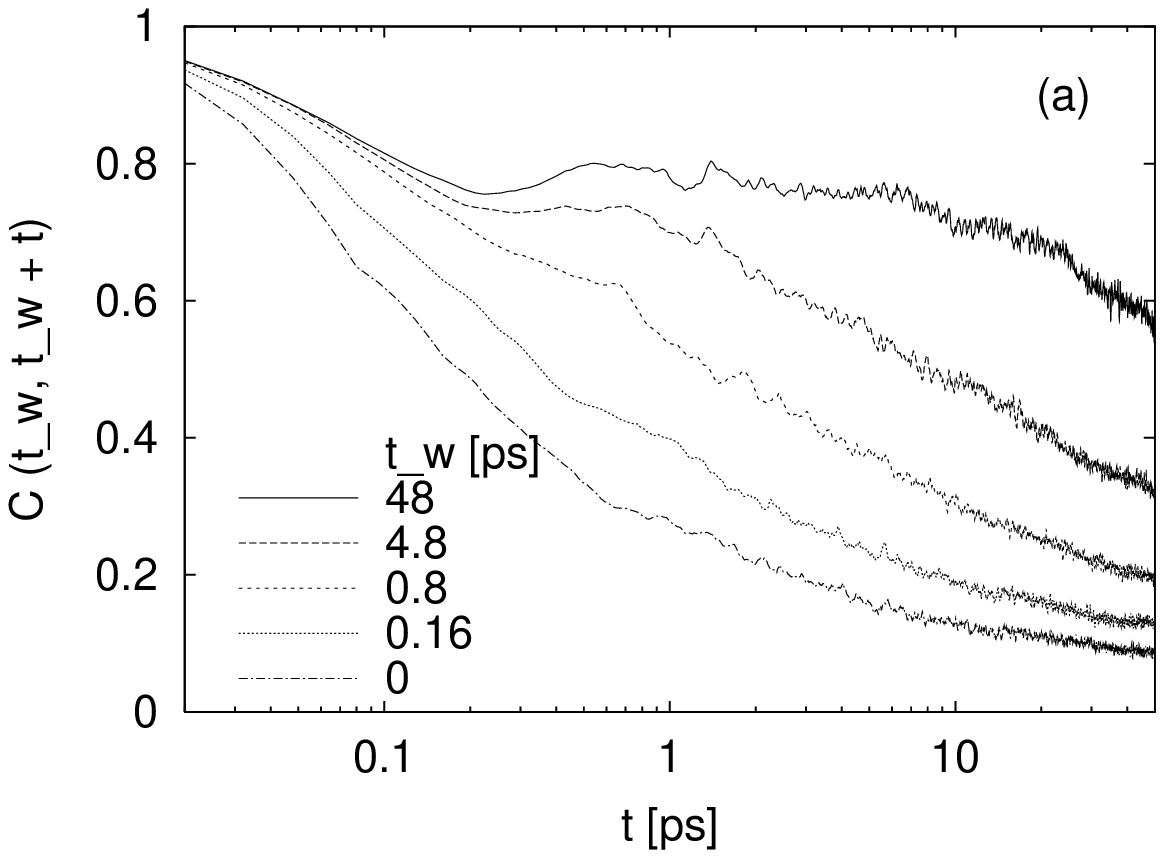,width=9cm}
    \epsfig{figure=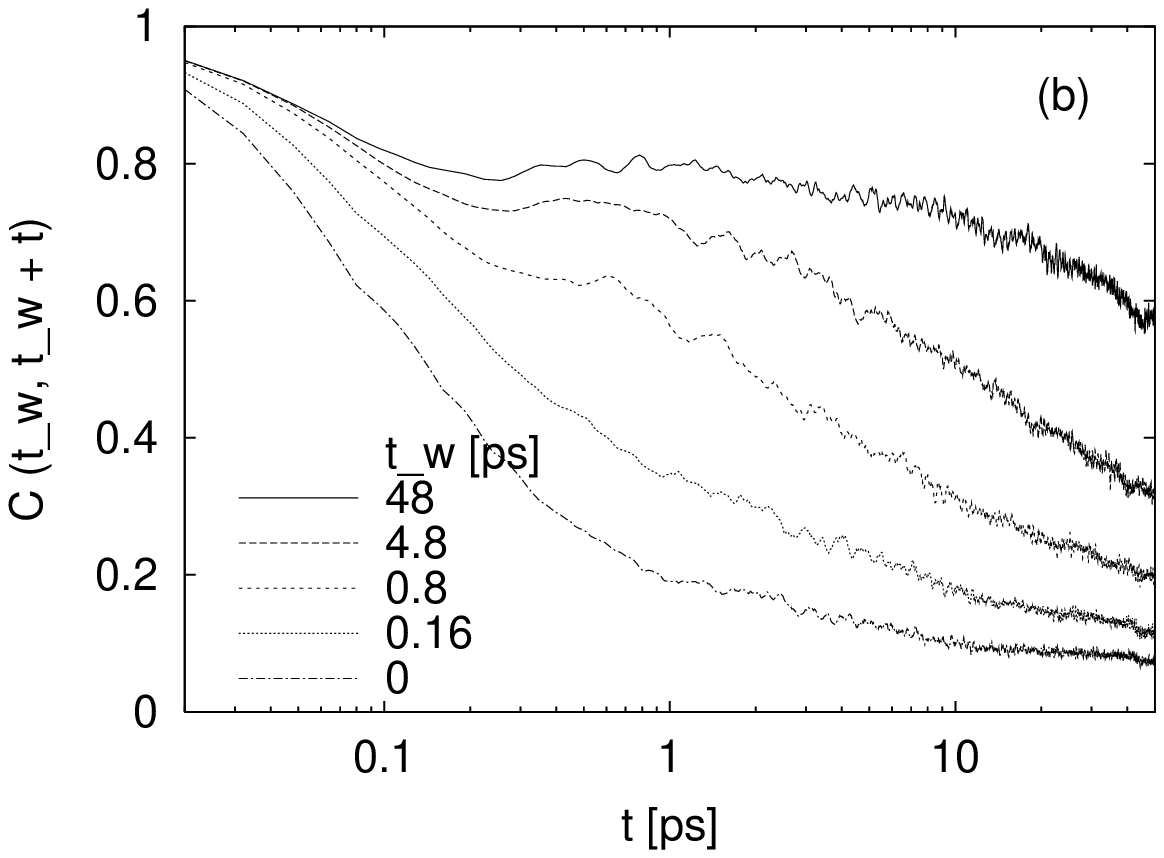,width=9cm}
\end{figure}

In recent simulations of this model, Kob and Barrat \cite{prl78-4581}
found a power law for the potential energy with an exponent of
$-0.144$, which is of the same order of magnitude as our result for
silica. We note, however, that in simulations for a soft sphere
model\cite{parisi}, thus with a potential similar to the Lennard-Jones
one, but using a Monte Carlo method instead of molecular dynamics, the
different value 0.7 has been found for the exponent; Kob and Barrat
mention the possibility that the disagreement is not caused mainly by
the comparatively small differences of the models, but by an influence
of the microscopic dynamics on the aging process. The fact that we
have obtained a result similar to theirs by using the same kind of
simulation, but a different potential, might be a further indication
for the validity of this assumption.

\subsection{Scattering function}

Figure \ref{fig:gsf} shows two sample plots for the time development
of the scattering function. A longer waiting time $\tw$ means that the
system has already had more time to relax after the quench than in the
case of a short one; the speed of the decorrelation therefore
decreases with growing $\tw$, so that the curves for larger values of
$\tw$ decay more slowly than those for short waiting times.

For short times the particles move ballistically, for long times they
show a diffusive motion due to collisions with each other;
particularly for low temperatures a plateau becomes visible between
these two ranges, which is symptomatic for the situation that
particles are trapped in ``cages'' formed by their neighbors and need
a comparatively long time to escape from these cages. The temperatures
in our simulations were not low enough to make this behavior visible
very clearly, but the tendency to form a plateau is recognizable in
the uppermost curves of figure \ref{fig:gsf}.

\begin{figure}[htb]
  \caption{Data for the same quenches as in figure \ref{fig:gsf},
    rescaled to have all curves fall together at $C = 0.5$.}
  \label{fig:scale}
  \epsfig{figure=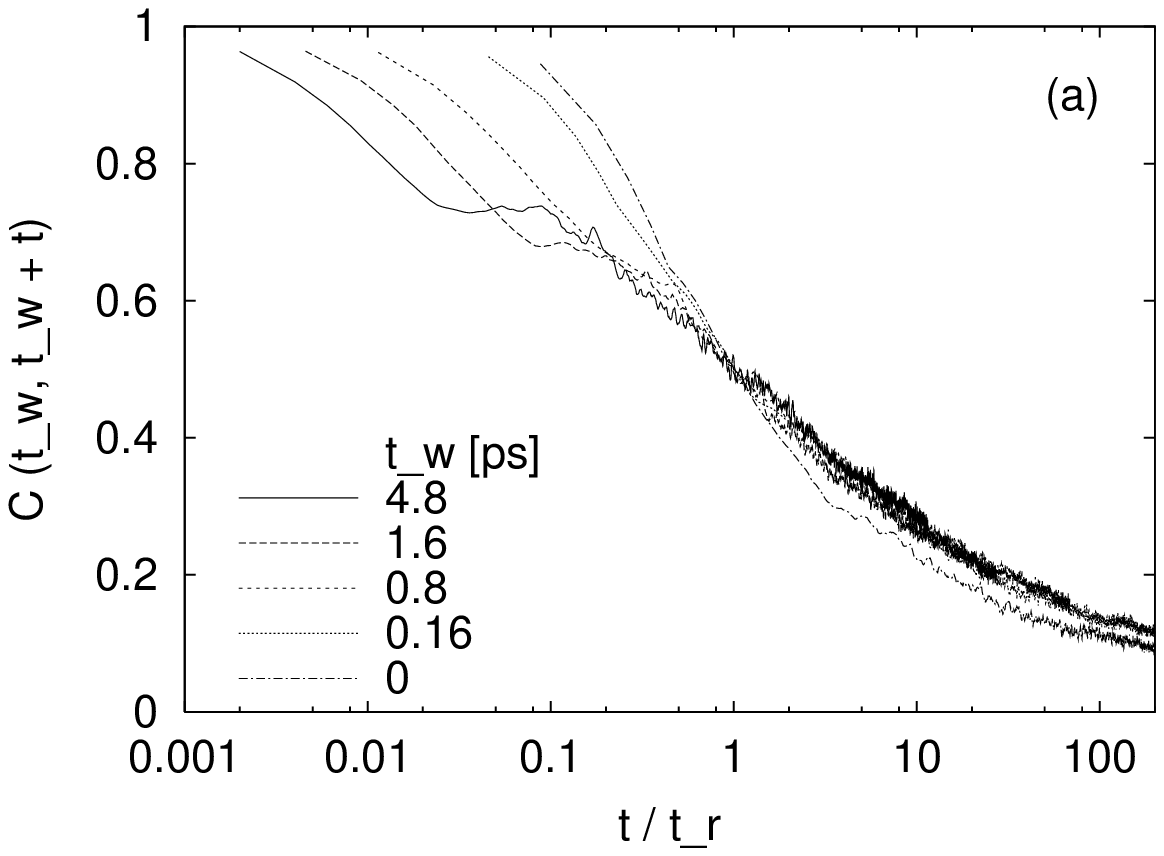,width=9cm}
  \epsfig{figure=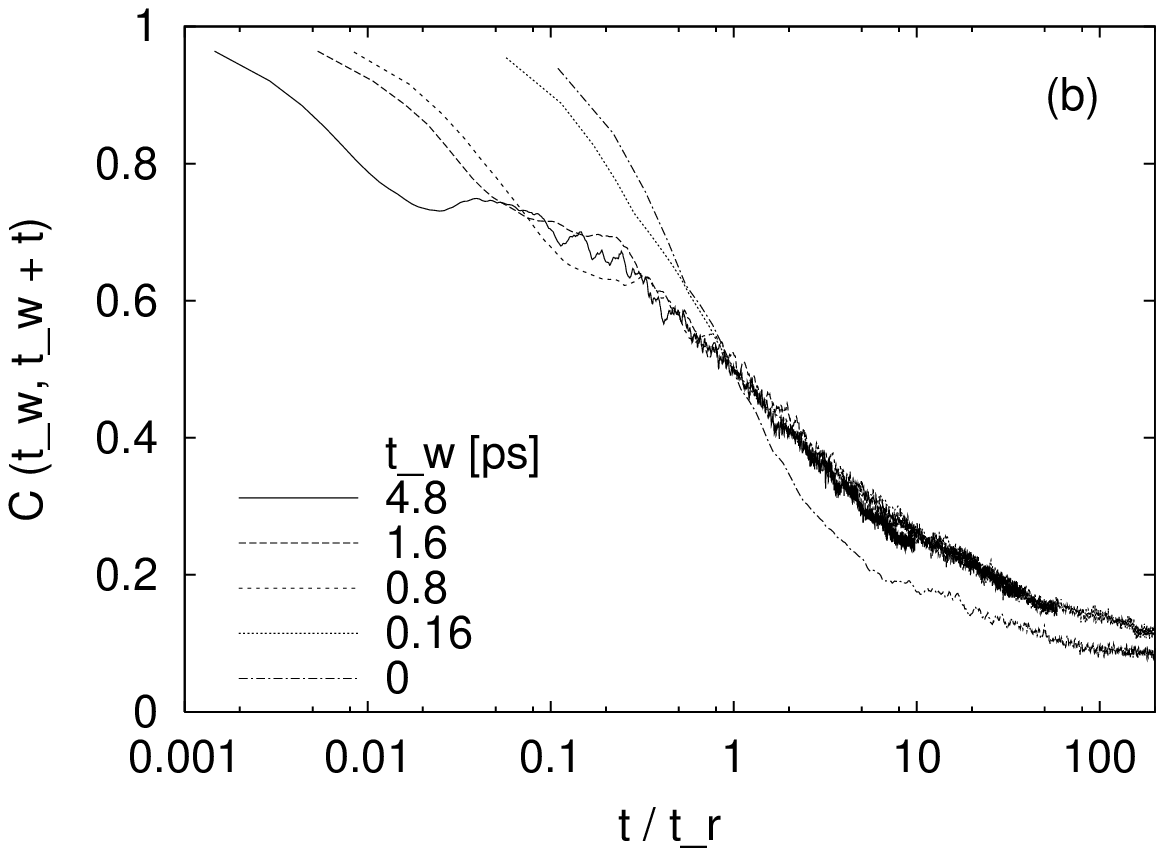,width=9cm}
\end{figure}

In passing we mention that the curves for longer waiting times in
figure \ref{fig:gsf} also show another interesting property, a dip
before the plateau or the final decay is reached, at a time of around
0.2~ps. This dip is related to the so-called boson peak, a dynamical
feature at around 1 THz ($10^{12}$ sec$^{-1}$) in the spectra of many
strong glass formers; several theoretical approaches have been
proposed to explain this peak, but until now no consensus has been
reached.

A natural scaling ansatz for two-time correlation functions
is 
\[
C (\tw, \tw + t) = f_{\rm stat} (t) + g_{\rm age} (t / \tr)\;,
\]
where the first term is the stationary part
\[
f_{\rm stat} (t) = \lim_{\tw\to\infty} C (\tw, \tw + t)
\]
and $g_{\rm age} (x)$ is a scaling function for the aging part, which
depends only on the ratio between the time $t$ and an effective
relaxation time $\tr$ that varies monotonically with the waiting time
$\tw$. In simple coarsening systems the off-equilibrium properties are
governed by a single length scale, the domain size or correlation
length $\xi$ that grows with the waiting time as $\xi \propto
\tw^\beta$, and $\tr$ is simply proportional to $\tw$.  However, in
aging experiments on glassy systems like polymer glasses \cite{struik}
and various spin glasses \cite{agesg} it was found that $\tr \propto
\tw^\alpha$ describes the data reasonably well when the exponent
$\alpha$ is used as a fit parameter (usually $\alpha$ is close to but
significantly different from 1).

To examine whether a similar statement can be made for silica, we
defined the relaxation time $t_{\rm r}$ of the scattering function as
the time necessary to reach the value $C = 0.5$ and checked the
scaling ansatz by plotting $C$ over $t / t_{\rm r}$. Figure
\ref{fig:scale} shows the results for two of the simulated quenches:
except for the case of small waiting times, the curves fall together
well for $t / t_{\rm r} \geq 1$, confirming the ansatz.

\begin{figure}[thb]
  \caption{Relaxation time $t_{\rm r}$ as a function of the waiting
    time $\tw$ for the quenches with initial temperature 6\,000~K.
    The dotted lines represent the approximation by a power law for
    long waiting times (from 0.8~ps upwards) with the exponent
    $\alpha$ as given in table \ref{tab:relax}.}
  \label{fig:relax}
  \epsfig{figure=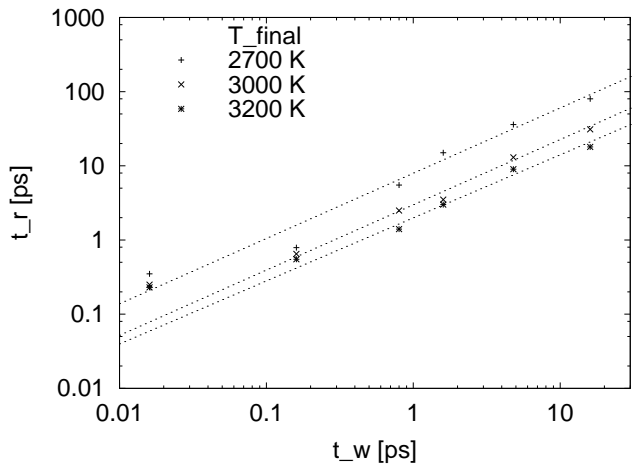,width=9cm}
\end{figure}

Plotting the relaxation time as a function of the waiting time reveals
that, again with the exception of small waiting times, they can be
approximated by a power law as well, as figure \ref{fig:relax}
demonstrates for the quenches with an initial temperature of 6\,000~K.
The exponents are listed in table \ref{tab:relax}; the average value
turns out to be $\alpha = 0.89$.

\begin{table}[bht]
    \caption{Estimated exponents $\alpha$ for the waiting time
      dependence of the relaxation time for the various combinations
      of initial (rows) and final (columns) temperatures. (Due to an
      error when the simulations were run, not enough data could be
      gathered to give reliable results for the quench from 8\,000 to
      3\,200~K.)}
    \begin{tabular}{@{\hspace{\tabcolsep}\extracolsep{\fill}}c|ccc}
& 2\,700~K & 3\,000~K & 3\,200~K \\
\hline
6\,000~K & 0.88 & 0.88 & 0.85 \\
7\,000~K & 0.86 & 0.95 & 0.77 \\
8\,000~K & 0.97 & 0.96 & ---
    \end{tabular}
    \label{tab:relax}
\end{table}

\begin{figure}
  \caption{Scaling plot of the scattering function (for $q = 7.2$) for
    a Lennard-Jones system with 8000 particles quenched from $T = 5$
    to $T = 0.35$ (glass transition at $T=0.46)$. {\bf Top}: The data
    are scaled according to a $t/\tr$ aging scenario, as in figure
    \ref{fig:scale} for the silica system. The data collapse is not
    very good. {\bf Bottom}: The data are scaled according to a
    $\ln((t+\tw)/\tau)/\ln(t_w/\tau)$ aging scenario \cite{prl81-930},
    the data collapse appears to be better.}
  \label{fig:lj}
  \epsfig{figure=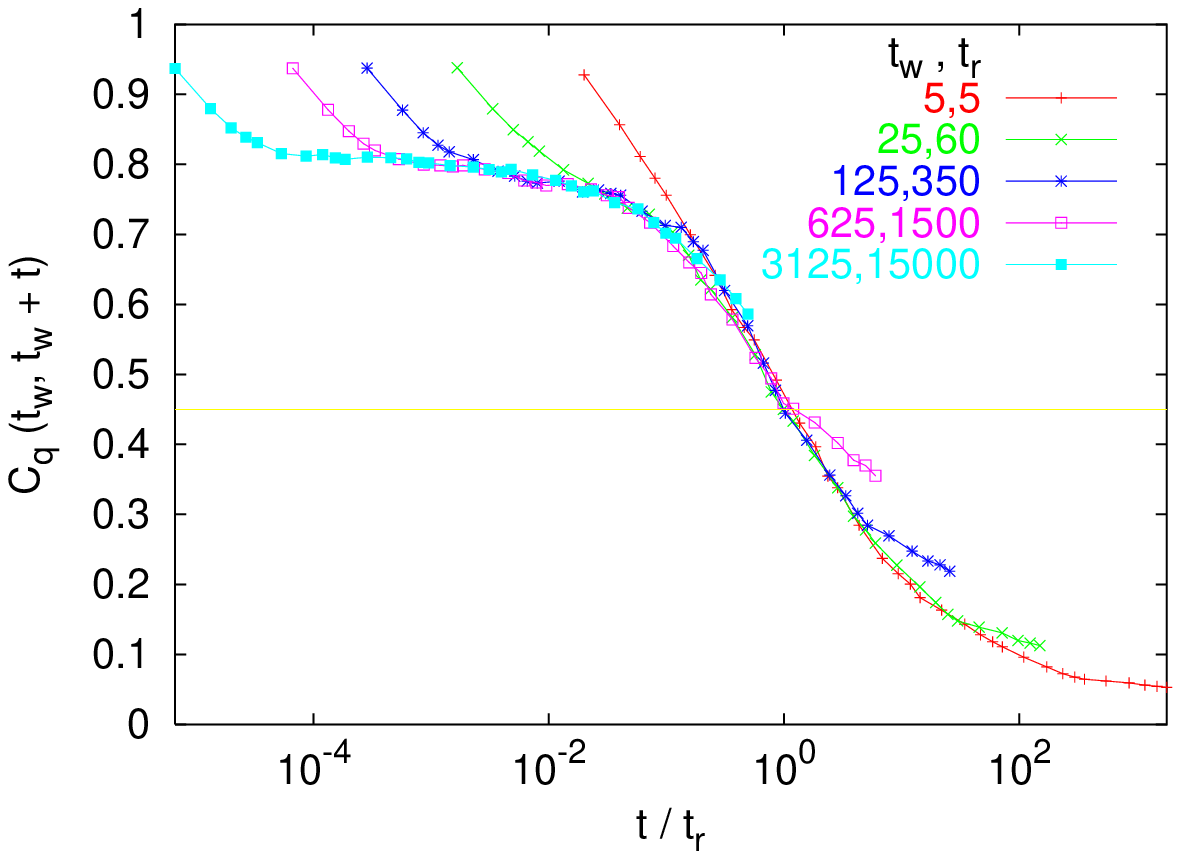,width=9cm}
  \epsfig{figure=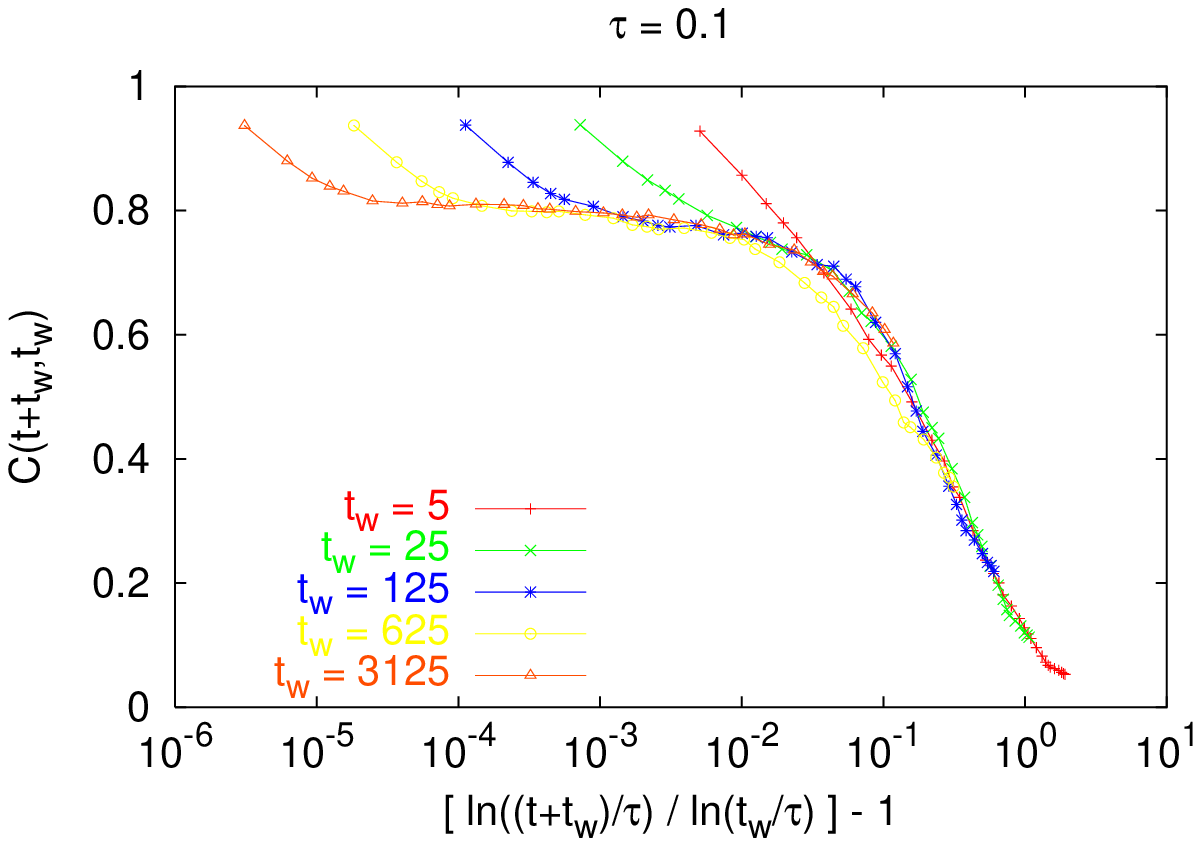,width=9cm}
\end{figure}

Again we compare our data to the one obtained in simulations of the LJ
model. Recent results there \cite{prl78-4581,prl81-930} present a somewhat
controversial picture of the behavior of the scattering function,
which we summarize in figure \ref{fig:lj}.

Here the data collapse is slightly better when using an
{\em activated dynamics} ansatz \cite{prl81-930}
\[
C (\tw, \tw + t) = f_{\rm stat} (t) + 
g_{\rm age} \left( \frac{\ln((t+\tw)/\tau)}{\ln(\tw/\tau)}\right)\;,
\]
although for higher temperatures the $t/t_r$ scaling scenario also
works fine with $\tr \sim \tw^\alpha$ and $\alpha \approx 0.88$
\cite{prl78-4581}, surprisingly similar to our result for
silica reported above.

To conclude we reported results of molecular dynamics simulations of a
model for a strong glass former, amorphous silica, and showed 1) that
the system is clearly aging on time scales of pico-seconds and 2) that
the off-equilibrium or aging part of the generalized structure
function obeys a $t/\tr$ scaling scenario with the effective
relaxation time (or effective age) obeying $\tr\propto\tw^{\alpha}$
with $\alpha \approx 0.88$. These results are in good agreement with
those reported recently for a fragile glass former, a binary
Lennard-Jones mixture.

\section*{Acknowledgement}
We thank Uwe M\"ussel for providing us with the data for figure
\ref{fig:lj} and Barbara Coluzzi for many stimulating discussions.
H.R.\ is grateful to the German research foundation for financial
support within a JSPS-DFG binational cooperation.

\end{document}